\documentstyle[12pt]{article}
\begin{document}
\title{The Schr\"{o}dinger Problem, L\'{e}vy Processes \\
and All That Noise \\
in Relativistic Quantum Mechanics}

\author{Piotr Garbaczewski\thanks{Kosciuszko Foundation Fellow.
Permanent address: Institute of Theoretical Physics, University of
Wroc{\l}aw, PL-50 204  Wroc{\l}aw, Poland}\, \,   and John R.
Klauder\thanks{Also Department of Mathematics} \\
Department of Physics, University of Florida, Gainesville, FL 32611\\
\and
and\\
Robert Olkiewicz\thanks{Supported by the KBN research grant No 2 P302
057 07}\\
Institute of Theoretical Physics, University of Wroc{\l}aw,\\
PL-50 204 Wroc{\l}aw, Poland}

\maketitle
\begin{abstract}
The main  purpose of the paper is an essentially probabilistic analysis
of relativistic quantum mechanics. It is based on the assumption that
whenever probability distributions arise, there
exists a  stochastic process that is either responsible for temporal
evolution of a given measure or preserves the measure in the stationary
case.
Our departure point is the so-called Schr\"{o}dinger  problem of
probabilistic evolution, which provides for a unique Markov stochastic
interpolation
between any given pair of boundary probability densities for a process
covering a fixed, finite duration of time, provided we have decided a
priori what kind of primordial dynamical semigroup transition
mechanism is involved.
In the nonrelativistic theory, including quantum mechanics,
Feyman-Kac-like kernels are the building blocks for  suitable
transition probability densities
of the process. In the standard "free" case (Feynman-Kac potential
equal to zero) the familiar Wiener noise is recovered.

In the framework of the Schr\"{o}dinger problem, the "free noise"
can also
be extended to any infinitely divisible probability law, as covered
by the L\'{e}vy-Khintchine formula. Since the  relativistic
Hamiltonians $|\nabla |$
and $\sqrt {-\triangle +m^2}-m$ are known to generate such laws,
we focus on them for the analysis of probabilistic phenomena, which
are shown to  be
associated with the relativistic wave (D'Alembert)  and matter-wave
(Klein-Gordon) equations, respectively.

We show that such stochastic processes exist and are spatial jump
processes.
In general, in the presence of external potentials, they  do not share
the  Markov property, except for stationary situations. A concrete
example of
the pseudodifferential Cauchy-Schr\"{o}dinger evolution is analyzed
in detail.
The relativistic covariance of related wave equations is exploited to
demonstrate how the associated stochastic jump processes comply with
the principles of special relativity.
\end{abstract}

\section{The analytic continuation in time of holomorphic semigroups
as a mapping between two families of stochastic processes}

\subsection{Gaussian exercises}
The Schr\"{o}dinger equation and the generalized heat equation are
connected by analytic continuation in time. The link (casually viewed as
a kind of analogy or correspondence) can be implemented by a rotation in
the complex time plane,
taking the Feynman-Kac kernel into the Green
function of the corresponding quantum mechanical problem, which is an
exploitation of properties of  holomorphic
semigroups generated by Laplacians and their sums with appropriate
potentials.

For $V=V(x), x\in R$, bounded from below, the generator $H=-2mD^2
\triangle + V$ is essentially selfadjoint on a natural dense subset of
$L^2
$, and the kernel $k(x,s,y,t)=[exp[-(t-s)H]](x,y)$ of the related
dynamical semigroup is strictly positive. The quantum unitary dynamics
$exp(-iHt)$ is a final result of the analytic continuation.

As repeatedly emphasized \cite{blanch,olk,nel}, any temporal evolution that
is analyzable in terms of a probability measure may be interpreted as
a stochastic process.  In view of the Born statistical interpretation
postulate for quantum mechanics, the analytic continuation in time
discussed
above induces a class of probability measures, namely, consider
$\rho (x,t)=|\psi (x,t)|^2 $ as the
density of a probability measure associated with a given solution $\psi
(x,t)$ of the Schr\"{o}dinger equation. Then, it is  possible to address
the problem of that  stochastic  dynamics which would be either
(i) measure
preserving or (ii) induce the  time evolution of the measure proper.
Keep in mind that the Schr\"{o}dinger equation itself \it is not \rm a
genuine partial differential equation of probability theory; rather it
is the Born postulate which embeds the unitary evolution problem into
the probabilistic framework.

A simple illustration of the analytic continuation in time is provided
by
considering the force-free propagation, where  the formal recipe gives
rise to the equations of motion  (one should be aware that to
execute a mapping for concrete
solutions, the proper  adjustment of the time interval boundaries is
indispensable):
$$i\partial _t\psi = -D\triangle \psi  \longrightarrow \partial _t
 \theta _*
=D\triangle \theta _*,$$
$${i\partial _t{\overline {\psi }} =D\triangle {\overline {\psi }}
\longrightarrow
 \partial _t\theta =-D\triangle  \theta ,} \eqno (1)$$
$$it\, \rightarrow \, t.$$
Then
$$\psi (x,t)= [\rho ^{1/2}exp(iS)](x,t) = \int dx'G(x-x',t)
\psi (x',0),$$
$${G(x-x',t) = (4\pi iDt)^{-1/2} \, exp[ -\, {(x-x')^2\over
 4iDt}],}\eqno (2)$$
$$\theta _*(x,t)= \int dx'k(x-x',t)\theta _*(x',0),$$
$$k(x-x',t)=(4\pi Dt)^{1/2}\, exp[-\, {(x-x')^2\over 4Dt}],$$
where the imaginary time substitution
$$k(x-x',it)=G(x-x',t)$$ $${k(x-x',t)=G(x-x',-it)}\eqno (3)$$
seems to persuasively suggest the notion of "evolution in imaginary
time",
which in the usual interpretation relates quantum theory to an
"imaginary
time diffusion". Here we shall emphasize a  different viewpoint in
which the quantum dynamics and the so-called Euclidean dynamics
\cite{zambr} (dealing with
the Wiener process and conditional Brownian motions, for example)
may both be seen as real time diffusions.

At this point let us observe that given the initial data
$${\psi (x,0)=(\pi \alpha ^2)^{-1/4}\, exp\, (-\, {x^2\over 2\alpha
^2})}\eqno (4)$$
the free Schr\" {o}dinger evolution  $i\partial _t \psi=-D\triangle
\psi $ implies that
$${\psi (x,t)= ({\alpha ^2\over \pi })(\alpha ^2 +2iDt)^{-1/2} \, exp\,
[\, -\,
{x^2\over 2(\alpha ^2+2iDt)}]}\eqno (5)$$
with
$$\rho (x,t)=|\psi (x,t)|^2 = {\alpha \over [\pi (\alpha ^4+
4D^2t^2)]^{1/2}}\,
exp(\, -\, {x^2\alpha ^2\over \alpha ^4 + 4D^2t^2})$$
$${=\int p(y,0,x,t)\rho (y,0)dy}\eqno (6)$$
$$p(y,0,x,t)=(4\pi Dt)^{-1/2} exp[\, -\, {(x-y-{2D\over \alpha ^2}yt)^2
\over 4Dt}]$$
where $p(y,0,x,t)$ is the (distorted Brownian) transition probability
density for Nelson' s diffusion \cite{nel,olk}. The transition
density  $p(y,0,x,t)$
 is not  uniquely specified by (6), but the nonuniqueness problem for
 the diffusion process involved may be resolved when we consider
 transition densities for arbitrary intermediate times.

For example \cite{olk}, if for convenience we set  $\alpha ^2=2,\,
D=1$, the transition density for two arbitrary times, $t>s$, reads
$$p(y,s,x,t) = [4\pi (t-s)]^{-1/2} exp[-{(x-cy)^2\over {4(t-s)}}],$$
$${c=c(s,t)=[{{(1-t)^2 + 2s}\over {1 + s^2}}]^{1/2},}\eqno (7)$$
$$\rho (x,t) = \int dy p(y,s,x,t) \rho (y,s),$$
Notice that the $s\downarrow  0$ limit of the transition density (7)
does not
coincide with  the rescaled form of (6), $|1-t|$ instead of $(1-t)$
appears in the exponent. It reflects the nonuniqueness  of the
definition of the
transition probability density as long as we do not  insist on  having
defined all intermediate densities as well \cite{olk}. Anyway, while
integrated with $\rho (x,0)$ they  give the same output at time $t>0$:
$${\rho _0(x)=(2\pi )^{-1/2} exp[-{x^2\over 2}] \longrightarrow
\rho (x,t)
= [2\pi (1+t^2)]^{-1/2} exp[- {x^2\over {2(1+t^2)}}]}\eqno (8)$$
Clearly  $\rho (x,t)$ admits a factorization  $\rho (x,t) =
|\psi (x,t)|^2 = \Theta (x,t) \Theta _*(x,t)$ . The
standard Madelung exponents
$R(x,t),S(x,t)$ such that $\psi (x,t) = exp[R(x,t)+iS(x,t)]$ are
given by
$${R(x,t)=-{1\over 4}ln2\pi (1+t^2) - {x^2\over {4(1+t^2)}},}
\eqno (9)$$
$$S(x,t)={x^2\over 4}{t\over {1+t^2}} - {1\over 2} arctan\,  (t)\, ,$$
and allow us to define the real functions $\Theta (x,t)=exp[R(x,t)
+S(x,t)] ,
\,  \Theta _*(x,t)= exp[R(x,t)-S(x,t)]$ which solve the pair of time
adjoint generalized diffusion equations
$$\partial _t\Theta=-\triangle \Theta + Q\Theta ,$$
$${\partial _t\Theta _*=\triangle \Theta _* - Q\Theta _*,}\eqno (10)$$
$$Q(x,t)=2{\triangle \rho ^{1/2}\over \rho ^{1/2}}.$$
The diffusion governed by the pair of adjoint equations (10) belongs to
the  category of "Nelson's diffusions" \cite{nel}, but only in the
framework presented here  can it be singled out uniquely. It is exactly
due to the "Schr\"{o}dinger problem" uniqueness theorem \cite{jam}.
It is really amazing that Schr\"{o}dinger originated the problem of
a stochastic interpolation between the prescribed input-output
statistics data long before the modern  probability theory was created.

An interesting observation is that we can give the transition density
(7) another form \cite{olk,zambr}
$${p(y,s,x,t)=k(y,s,x,t){\Theta (x,t)\over {\Theta (y,s)}},}\eqno (11)$$
$$lim_{\triangle s\downarrow 0}{1\over {\triangle s}}[1 - \int
k(y,s,x,s+ \triangle s)dx] = Q(y,s)$$

On the other hand, coming back to the previous notation and (5), we can
straightforwardly pass  to
$${\theta _*(x,t) := \psi (x,-it) = ({\alpha ^2\over \pi })^{1/4}
 (\alpha ^2 +2Dt)^{-1/2}
exp\, [-\, {x^2\over 2(\alpha ^2+ 2Dt)}]}\eqno (12)$$
Let us confine $t$ to the time interval $[-T/2,T/2]$ with $DT<\alpha ^2$.
Then we
arrive at
$$\partial _t \theta _* =D\triangle \theta _*,$$
$$\partial _t\theta = -D\triangle \theta
 ,$$
$${-{T\over 2}\leq t\leq {T\over 2},}\eqno (13)$$
$$\theta (x,t)=({\alpha ^2\over \pi })^{1/4} (\alpha ^2 - 2Dt)^{-1/2} \,
 exp
[\, -\, {x^2\over 2(\alpha ^2 - 2Dt)}]$$
where (we use an overbar to distinguish between the probability densities
(14) and (6) or (7) respectively; notice also that $\theta $ replaces
$\Theta $)
$${\overline {\rho }(x,t) = \theta (x,t)\theta _*(x,t) =
[{\alpha ^2\over \pi (\alpha ^4 -4D^2t^2)}]^{1/2}\, exp[\, -\,
{{\alpha ^2x^2}\over {\alpha ^4 - 4D^2t^2}}]}\eqno (14)$$
with the interesting and certainly unexpected--if one  follows
traditional Brownian intuitions--outcome that:
$${{\overline {\rho }}(x, - t)={\overline {\rho }}(x,t)}\eqno (15)$$
for all $|t|\leq T/2$. The density (14) refers to a conditional Brownian
motion, and
the interpolating probability density  can be represented as
the conditional probability density (identifiable as the Bernstein
density \cite{zambr})
$${\overline {\rho }}(x,t) = {{\overline {k}(0,-\alpha _0,x,t)
\overline {k} (x,t,0,\alpha  _0)}
\over {\overline {k}(0,-\alpha _0,0,\alpha _0)}}, $$
$${\overline {k}(y,s,x,t)\, :=\, [4\pi D(t-s)]^{-1/2} exp[ -{{(x-y)^2}
\over
{4D(t-s)}}],}\eqno (16)$$
$$\alpha _0 = {\alpha ^2\over {2D}}$$
Since ${\overline {\rho  }}(x,t)$ trivially factorizes into a product
of solutions of time adjoint heat equations
(set $\theta (x,t) = [\overline {k}(0,-
\alpha _0,0,\alpha _0)]^{-1/2} \overline {k}(x,t,0,\alpha _0)$), we
have in hand the microscopic transport recipe \cite{zambr}
$${p(y,s,x,t)=\overline {k}(y,s,x,t){{\theta (x,t)}
\over {\theta (y,s)}}}
\eqno (17)$$
with the heat kernel $\overline {k}(y,s,x,t)$, (16), and the property
(to be compared with (11)):
$${lim_{\triangle s\downarrow 0}{1\over {\triangle s}} [1 -
\int \overline {k}(y,s,x,s+\triangle s)dx] = 0 }\eqno (18)$$

The resemblance of formulas (11) and (17), (18) is not accidental, and
suggests that any given Feynman-Kac kernel can be used to generate a
probability measure.
In addition, we should keep in mind that the two levels of probabilistic
description i.e. (11) and (17), are \it indirectly \rm linked by the
analytic continuation in time of a holomorphic semigroup with the
Laplacian as its generator.

{\bf Remark 1}: The time developement of a density $\rho (x,t)$
($\overline {\rho }$ respectively)  is dictated by the Fokker-Planck
(second Kolmogorov in the probabilistic lore) equation. It is instructive
to notice that $i\partial _t\psi = -\triangle \psi $ upon setting
$v=2Re {{\nabla \psi }\over \psi}$ and $u=2Im {{\nabla \psi }
\over \psi }$
gives rise to
$\partial _t\rho =-\nabla (v\rho )$, which may be rewritten  as
$\partial _t\rho =\triangle \rho - \nabla (b\rho ) $ with $b=u+v$.
Proceeding analogously with $\overline {\rho } =k(x_1,t_1,x,t)
k(x,t,x_2,t_2)/k(x_1,t_1,x_2,t_2)$ where $k(y,s,x,t)$ is the heat kernel,
and $s<t,\, t_1<t<t_2$, while $t_1,x_1,t_2,x_2$ are fixed, we immediately
recover $\partial _t\overline {\rho }=\triangle \overline {\rho} -
\nabla (b \overline {\rho })$ with $b=b(x,t)=2\nabla k(x,t,x_2,t_2)/ \\
k(x,t,x_2,t_2)$.

Furthermore to this remark, let us emphasize that the emergence of the
nonvanishing
drift field $b(x,t)$ is not connected with any external force .
It is a traditional assumption when studying the Brownian motion in a
conservative force field to define the drift as being  proportional to
the force itself (Stokes law); these Smoluchowski diffusions form a
subclass of problems we are considering \cite{blanch}, however the
previous drift is an exclusive effect of the conditioning.

\subsection{The Schr\"{o}dinger problem: from Feynman-Kac kernels to
probability measures}

The previous examples are very particular solutions of what we call
\cite{blanch,olk} \it the Schr\"{o}- dinger problem \rm of deducing the
probabilistic interpolation (stochastic process) consistent with a
given pair
of boundary measure data at fixed initial and terminal time instants
$t_1 <t_2$. Originated by Schr\"{o}dinger himself \cite{schr}, the
problem
was solved much later \cite{jam,zambr} by invoking the machinery of
Bernstein stochastic processes; see also Ref. 4. For our purposes
the relevant
information is that \cite{jam}, if the interpolating process is to
display the Markov
property,  then it has to be specified by the joint probability measure
($A$ and $B$ are Borel sets in $R$):
$$m(A,B)= \int_A dx\int_B dy \, m(x,y),$$
$${\int_R m(x,y) dy = \rho (x,t_1),}\eqno (19)$$
$$\int_R m(x,y) dx = \rho (y,t_2),$$
where we assign densities to all measures to be dealt with, and the
density $m(x,y)$ is given  in the functional form
$${m(x,y) = f(x)k(x,t_1,y,t_2)g(y)}\eqno (20)$$
involving  two unknown  functions $f(x)$ and $g(y)$ which are of the same
sign and
nonzero, while $k(x,s,y,t)$ is any bounded strictly positive
(dynamical semigroup)  kernel defined for all times
$t_1\leq s<t\leq t_2$.
The integral equations (19) determine  functions $f(x),g(y)$ uniquely
(up to constant factors) in  this case \cite{jam}.

By denoting $\theta _*(x,t)=\int f(z) k(t_1,z,x,t) dz $ and
$\theta (x,t)=
\int k(x,t,z,t_2)g(z) dz$ it follows \cite{jam,zambr,blanch,olk} that
$${\overline {\rho}(x,t) = \theta (x,t)\theta _*(x,t) = \int p(y,s,x,t)
\overline {\rho }(y,s) dy ,}\eqno(21)$$
$$p(y,s,x,t) = k(y,s,x,t){{\theta (x,t)}\over \theta (y,s)},$$
$$t_1\leq s<t\leq t_2$$
Hence the transition densities are
intrinsically entangled with the dynamical semigroup kernels in the
solution
of the Schr\"{o}dinger stochastic interpolation problem. The crucial
step in
the construction of any explicit propagation consistent with the
boundary measure data is to decide what is the appropriate  dynamical
semigroup.

We shall address the issue in its full generality. Strictly positive
semigroup
kernels generated by Laplacians plus suitable potentials are very special
examples in
a surprisingly rich encompassing family. First of all, the concept of the
"free noise", normally characterized by a Gaussian probability
distribution
appropriate to a Wiener process, can be extended to all infinitely
divisible
probability distributions via the L\'{e}vy-Khintchine formula
\cite{bre,sim}.
It expands our framework from continuous diffusion processes to  jump
or combined diffusion--jump propagation scenarios. All such (L\'{e}vy)
processes
are associated with strictly positive dynamical semigroup kernels, and
all of them give rise to Markov solutions of the Schr\"{o}dinger
stochastic interpolation problem (19)-(21).

{\bf Remark 2}: Apart from the wealth of physical phenomena described
in terms
of Gaussian stochastic processes, there is a number of physical
problems where
the Gaussian tool-box proves to be insufficient to provide satisfactory
probabilistic explanations. Non--Gaussian L\'{e}vy processes
naturally
appear in the study of transient random walks when long-tailed
distributions
arise \cite{west,wer,fog}. They are also found necessary to analyze fractal
random walks
\cite{mand}, intermittency phenomena, anomalous diffusions, and
turbulence at high Reynolds numbers \cite {west,wang,klaf}.

Let us consider  Hamiltonians of the form $H=F(\hat{p})$, where
$\hat{p}=-i
\nabla $ stands for the momentum operator and  for
$-\infty <k<+\infty $, $F=F(k)$ is a real valued,
bounded from below, locally integrable function. Then,
$exp(-tH)=\int_{-\infty }^ {+\infty } exp[-tF(k)] dE(k) $, $t\geq 0$,
where $dE(k)$ is the spectral measure of $\hat{p}$.

Most of our discussion  will pertain to processes in one spatial dimension,
and let us specialize the issue accordingly. Because  $(E(k)f)(x)=
{1\over {\sqrt {2\pi }}}\int_{-\infty }^{k} exp(ipx) \hat{f}(p) dp $,
where  $\hat{f}$
is the Fourier transform of $f$, we learn that
$$[exp(-tH)]f(x) = [\int_{-\infty }^{+\infty } exp(-tF(k)) dE(k)f](x)
=  $$
$${{1\over {\sqrt {2\pi }}}\int_{-\infty }^{+\infty }exp[-tF(k)]
{d\over {dk}}
[\int_{-\infty }^{k} exp(ipx) \hat{f}(p) dp]dk = }  \eqno (22)$$
$${1\over {\sqrt {2\pi }}}\int_{-\infty }^{+\infty } exp(-tF(k)) exp(ikx)
\hat{f}(k) dk = [exp(-tF(p)) \hat{f}(p)]^{\vee }(x)$$
where the superscript $\vee $ denotes the inverse Fourier transform.

Let us set $k_t={1\over {\sqrt {2\pi }}}[exp(-tF(p)]^{\vee }$, then the
action of $exp(-tH)$ can be given in terms of a convolution:
$exp(-tH)f = f*k_t$, where $(f*g)(x): =\int_R g(x-z)f(z)dz $.

 We shall restrict consideration only to those $F(p)$ which give rise to
 positivity preserving semigroups:  if $F(p)$ satisfies the celebrated
 L\'{e}vy-Khintchine formula, then $k_t$ is a positive measure for all
 $t\geq 0$.
The most general case refers to a contribution from three types of
processes:  deterministic, Gaussian, and an exclusively jump process. We
shall concentrate on the integral part of the L\'{e}vy-Khintchine formula,
which is responsible for arbitrary stochastic jump features:
$${F(p) = -  \int_{-\infty }^{+\infty } [exp(ipy) - 1 -
{ipy\over {1+y^2}}]
\nu (dy)}\eqno (23)$$
where $\nu (dy)$ stands for the so-called L\'{e}vy measure
\cite{bre,kolm}.

The disregarded Gaussian contribution would read $F(p)=p^2/2$; cf. Refs.
3-7 for an exhaustive discussion of  related topics. In this case we
know in
detail how the analytic continuation in time of the Laplacian generated
holomorpic semigroup induces a mapping to a quantum mechanical
(since the Schr\"{o}dinger equation is involved) diffusion processes.

Our further attention will focus on two selected choices  for the
characteristic exponent $F(p)$, namely:
$F_0(p)=|p|$ and $F_m(p) =\sqrt {p^2 + m^2} - m, m>0$, where we have
chosen suitable units so as to eliminate inessential parameters.
(The relativistic Hamiltonian is better known in the form
$\sqrt {m^2c^4+c^2p^2 }-mc^2$ where $c$ is the velocity of light.)

The respective Hamiltonians (semigroup generators)  $H_0, \, H_m$ are
pseudodifferential operators. The semigroup kernels $k^0_t,\, k^m_t$
in view
of the "free noise"restriction (no potentials, will be defined in
below) are transition densities of
the jump (L\'{e}vy) processes  regulated by the  corresponding  L\'{e}vy
measures $\nu _0(dy),\, \nu _m(dy)$.
The affiliated  Markov processes solving the Schr\"{o}dinger problem
(19)-(21) immediately follow.
It is instructive to notice that as in the case of Gaussian derivations
(1), it is \cite{blanch} the case  $\theta (x,t)\equiv  1,\,
\theta _*(x,t):= \overline {\rho }(x,t)$
for which the pseudodifferential analog of
the Fokker-Planck equation, as a consequence of
$[exp(-tH)\overline {\rho }](x)=\overline {\rho }(x,t)$ and in view of
the identification $F(p\rightarrow -i\nabla ):=H$ takes the
fundamental  form
$${F_0(p)\Longrightarrow \partial _t{\overline {\rho }}(x,t)=
 - |\nabla|{\overline {\rho }}(x,t)}\eqno (24)$$
 or
 $${F_m(p) \Longrightarrow \partial _t{\overline {\rho }}(x,t)= -
 [\sqrt {-
 \triangle  + m^2} - m]\overline {\rho }(x,t)}\eqno (25)$$
 respectively. Let us emphasize that the existence and uniqueness of
 solutions proof for the Schr\"{o}dinger problem  extends to all cases
 governed by the infinitely divisible probability laws, and has nothing
 to do with the "nonrelativistic" or  "relativistic" options. The
 particular choice of semigroup  generators, which are called
 "relativistic Hamiltonians" links the standard Schr\"{o}dinger problem
 discussion with relativistic dynamics. But \it only \rm after  an
 analytic  continuation in time, unless only stationary problems are
 studied (see the forthcoming discussion).

Although the pseudodifferential generator of the semigroup implies that
the
Fokker-Planck equation is no longer exclusively differential but an
integro-differential equation, each solution $\overline {\rho }(x,t)$
in  the
present case is nevertheless a solution of a partial differential
equation of higher order.
 Specifically, the respective partial differential equations are of the
 second order:
$${F_0(p)\Longrightarrow \Box _E\overline {\rho }(x,t) = (\triangle _t +
\triangle )\overline {\rho }(x,t) = 0.}\eqno (26)$$
Alternatively, if we set
$\overline {\rho }(x,t):= \tilde{\rho }(x,t)\, exp(mt)$ in (25) then:
$${F_m(p)\Longrightarrow (\triangle _t + \triangle )\tilde {\rho }(x,t) =
m^2\tilde {\rho }(x,t)}$$
$${\Downarrow }\eqno (27)$$
$$( - \Box _E + m^2)\tilde {\rho }(x,t) = 0$$
where $\partial _t\tilde {\rho } = - \sqrt {-\triangle +m^2}
\tilde {\rho }$
holds true instead of (25).

Our two semigroups are holomorphic \cite{sim1}, hence we can replace
the time parameter
$t$ by a complex one $\sigma =t+is, \,t>0$  so that $exp(-\sigma H)=
\int_R
exp(-\sigma F(k))\, dE(k)$.  Its  action is defined
by
$${[exp(-\sigma H)]f = [(\hat{f}exp(-\sigma F)]^{\vee } = f*k_{\sigma }}
\eqno (28)$$
to be compared with (22). Here, the kernel reads $k_{\sigma }={1
\over {\sqrt {2\pi }}}[exp(-\sigma F)]^{\vee }$. Since $H$ is
selfadjoint, the
limit $t\downarrow 0$ leaves us with the unitary group $exp(-isH)$,
acting in
the same way: $[exp(-isH)]f = [\hat{f} exp(-isF)]^{\vee }$, except
that now
$k_{is}: = {1\over {\sqrt {2\pi}}}[exp(-isF)]^{\vee }$ in general is
\it not
\rm a measure. In view of  unitarity, the unit ball in $L^2$ is an
invariant of the dynamics. Hence density measures can be associated with
solutions
of the Schr\"{o}dinger pseudodiferential equations:
$${F_0(p)\Longrightarrow i\partial _t \psi (x,t) =  |\nabla | \psi (x,t)}
\eqno (29)$$
or
$${F_m(p)\Longrightarrow i\partial _t\psi (x,t) =
[\sqrt {-\triangle + m^2}
- m ]\psi (x,t)}\eqno (30)$$
provided with the appropriate initial data functions $\psi (x,0)$.

An obvious consequence of (29),(30) is that  the partial differential
equation of the second order (26) takes on a familiar \it relativistic
\rm form
$${F_0(p)\Longrightarrow  \Box \psi (x,t) := ( - \triangle  +
\triangle _t)
\psi (x,t) = 0}\eqno (31)$$
while after setting $\psi (x,t) = \tilde {\psi }(x,t)\, exp(imt)$, we
arrive
at the Klein-Gordon equation:
$${F_m(p)\Longrightarrow (\Box + m^2)\tilde {\psi }(x,t) = 0 }
\eqno (32)$$
where the D'Alembert operator $\Box = -\triangle +  \triangle _t $
replaces
its Euclidean counterpart  $-\Box _E$ in (27).

We have thus reached a point, at which  the main questions addressed in
the present paper can be precisely stated:

(i) What are the stochastic processes consistent with the probability
measure
dynamics $\rho (x,t)=|\psi (x,t)|^2$  determined by  pseudodifferential
equations (29) and (30)?

(ii) Can we extend the Schr\"{o}dinger problem idea to the special
relativistic domain and be able to reproduce the interpolating stochastic
process
from the given input $\rho (x,t_1)$ and output $\rho (x,t_2), \, t_1<t_2$
statistics data, just as in the nonrelativistic (Laplacian generated
motion) case ?

(iii) To what extent can we attribute a definite probabilistic meaning to
solutions of the relativistic wave equations (31),(32) ?

\section{Can we associate Feynman-Kac kernels with the pseudodifferential
Schr\"{o}dinger dynamics ?}

Given the Schr\"{o}dinger equations (29),(30). To set them in the \it
Schr\"{o}dinger problem  \rm framework of Section 1.2 we need to choose
any normalized solution and then take the associated probability density
$\overline {\rho }(x,t):=|\psi (x,t)|^2$ as the boundary data at times
$t_1<t_2$.  However, as stated before  some additional requirements must
be met, specifically the Markov property is necessary \cite{jam,zambr}.
One should keep in mind that if we do not insist on the Markov property for
the interpolating process, then a solution of the problem involves the
general
Bernstein processes \cite{jam} for which a reformulation in terms of a
pair of
time adjoint generalized diffusion equations no longer exists.

We have chosen two rather special pseudodifferential counterparts of the
Laplacian guided by two reasons: (i) their similarity on analytic grounds
(the same
criteria \cite{carm1} for the existence of the bound state spectrum if
summed with
suitable potentials, which we shall need in the sequel), (ii)
the claim of Ref.\cite{ang} that the pertinent stochastic process
in the mass $m>0$ case actually displays the Markov property.

If the Markov property  would hold true for the relativistic Hamiltonian
generated dynamics, we would be able to repeat almost all steps of the
previous
Schr\"{o}dinger problem analysis \cite{zambr,carm,blanch,olk}.
However, the
situation is not that simple, and the subsequent argument \it
excludes \rm the Markov property, in all nonstationary situations,
in a flat
contradiction with  general statements by De Angelis \cite{ang}.

Before embarking on this issue, let us introduce some probabilistic
notions, which will tell  us how to work with  pseudodifferential
operators.
We shall notice that for explicit computational purposes, the Cauchy
generator $|\nabla |$ is much more suited than the $m>0$ relativistic
Hamiltonian.
It is a real disadvantage when dealing with L\'{e}vy processes that
rather
limited number of concrete examples is available, in contrast to the
wealth of the general theory.

The L\'{e}vy-Khintchine formula  (23) tells us that the action of the
Hamiltonian $H=F(-i\nabla )$ on a function in its domain can be
represented
as follows  \cite{bre,kolm}):
$${(H\, f)(x)\, =\, - \int_R [f(x+y) - f(x) - {{y\, \nabla f(x)}
\over {1+y^2}
}]\, \nu (dy)}\eqno (33)$$

It is important to observe that for the "free noise" processes
whose semigroup generators are $|\nabla |$ and $\sqrt {-\triangle +m^2}
- m$
we do know explicitly their kernels (transition probability
densities) and the involved L\'{e}vy measures, as well as about the
extension
of the Feyman-Kac \it path integral \rm construction of the semigroup
kernels
to these particular L\'{e}vy processes \cite{carm1,ichi,ichi1}, in case of
arbitrary space dimensions.  Therefore we feel free to use the Feynman-Kac
kernel notion instead of the semigroup kernel.

For the Cauchy process, whose generator is $|\nabla |$, we deal with a
 probabilistic classics \cite{bre,kolm}:
$${\overline {\rho }(x,t) = {1\over {\pi }}\, {t\over {t^2 + x^2}}
\Longrightarrow  k^0(y,s,x,t) = {1\over {\pi }}{{t-s}\over {(t-s)^2 +
(x-y)^2}}]}\eqno (34)$$
$$0<s<t$$
$$\langle exp[ipX(t)]\rangle := \int_R exp(ipx) \overline {\rho }(x,t)\,
dx = exp[-tF_0(p)] = exp(-|p|t)$$
The characteristic function of $k^0(y,s,x,t)$ for $y,s$ fixed,  reads
$exp[ipy - |p|(t-s)]$, and the L\'{e}vy measure needed to evaluate the
L\'{e}vy-Khintchine integral reads \cite{fel,wat,ichi}:
$${\nu _0(dy): = lim_{t\downarrow 0} [{1\over t}k^0(0,0,y,t)]dy ={{dy}
\over
{\pi y^2}}}\eqno (35)$$

In the case of the relativistic generator $\sqrt {-\triangle +m^2} - m$,
formulas determining the  stochastic jump process  are  much less appealing
\cite{ichi,ichi1}:
$${\langle exp[ipX(t)]\rangle := exp[-tF_m(p)] = exp[-t(
\sqrt {p^2+m^2}-m)]}
$$
$${\overline {\rho }(x,t) = {m\over {\pi}} {texp(mt)
\over {\sqrt {x^2+t^2}}}
\, K_1(m\sqrt {x^2+t^2})}\eqno (36)$$
$$[exp(-(t-s)F_m(-i\nabla ))](x-y) = k^m(y,s,x,t) := \overline {\rho }
(x-y,
t-s)$$
$$\nu _m(dy) = {m\over {\pi |y|}} K_1(m|y|) dy$$
where $K_1(z)$ is the modified Bessel function of the third kind of
order 1.

We are interested  in  acting with the pseudodifferential generators
$H=F(-i
\nabla )$ on functions in the exponential form (recall the familiar
Madelung
procedure in the Gaussian case) $f(x,t) = exp \Phi (x,t)$:
$$(Hexp\Phi )(x) = - \int_R  [exp\Phi (x+y)\, - \, exp\Phi (x) \, -\,
{{y\Phi '(x)exp\Phi (x)}\over {1+y^2}}]\nu (dy)=$$
$$={exp\Phi (x)\, \int_R [exp(\Phi (x+y) - \Phi (x))\, -\, 1\, -\,
{{y\Phi '
(x)}\over {1+y^2}}]\nu (dy)}\eqno (37)$$
where $\Phi '(x)=\nabla \Phi (x)$.  Since $(H\Phi )(x) = -
\int_R [\Phi (x+y)
-\Phi (x) -y\Phi '(x)/(1+y^2)]\nu (dy)$, we can make a safe
rearrangement of (37):
$${(H\exp\Phi )(x) = exp\Phi (x)\, [(H\Phi )(x)\, - \, \int_R
(exp\Phi _{xy}
\, -\, 1\, - \Phi _{xy})\nu (dy)]}\eqno (38)$$
$$\Phi _{xy} := \Phi (x+y) - \Phi (x)$$

In application to the pseudodifferential dynamics
$i\partial _t\psi (x,t)=
(H\psi )(x,t)$ with $\psi =exp(R+iS)$, we shall  investigate its
implications
for the real functions $\Theta =exp(R+S)$ and $\Theta _*=exp(R-S)$;
our
argument will admit a trivial extension from $H$ to $H+V$ situations.

{\bf Remark 3}: Experience \cite{garb,blanch} with the Gaussian
(standard Laplacian generated) noise proves that the Madelung
substitution
$\psi (x,t) =exp [R(x,t)+iS(x,t)]$ would associate with the
Schr\"{o}dinger  equation
a pair of time adjoint generalised diffusion equations where the
Feynman-Kac potential (time dependent in the general case) equals
${1\over {2mD}}
[2Q(x,t) - V(x)]$. Here $Q(x,t)=2mD^2{{\triangle \rho ^{1/2}}
\over {\rho }
^{1/2}}(x,t)$  and $V(x)$ is taken as an external conservative force
potential. Let us emphasize that $V(x)$  actually \it was \rm  the
Feynman-Kac potential of the dynamical semigroup prior to the analytic
continuation in time
procedure. The mapping $V(x)\rightarrow 2Q(x,t)-V(x)$ is an effect of
the
analytic continuation in time, as manifested on the level of the
associated  Feynman-Kac kernels. Previous comments suggest that the
nonrelativistic
formalism can be viewed as a kind of probabilistic reinterpretation of
Bohm's point of view \cite{hol}. Specifically, the function
$-Q(x,t)$ is the familiar de Broglie-Bohm "quantum potential".
The analogous
connection is generally invalid in the context of the Klein-Gordon
equation, as explained in Ref.\cite{hol}.

To this remark, let us add
that the analytic  continuation in time as outlined in Section 1 is not
a well known procedure although repeatedly mentioned in the previous
publications \cite{blanch,olk}. It is new as a regular method. Also,
it is not  a part of Nelson's  "quantum fluctuations" \cite{nel} point of
view, just as the Feynman-Kac kernels were not ingredients of Nelson's
theory. The idea is new, though developed on the basis of Zambrini's
\cite{zambr} and Carmona's \cite{carm} research. Refs.2-6  give a
complete description of the state of art in this respect. Therefore we do
not propose to indicate any further connections with Nelson's stochastic
mechanics than by referring to Nelson's monograph \cite {nel}. The paper
itself proposes a new, self-contained probabilistic analysis which, in
addition to Zambrini's work is motivated by the paper due to De Angelis
\cite{ang}.

In view of (38), the pseudodifferential Schr\"{o}dinger equation
$i\partial _t
\psi (x,t)=H\psi (x,t)$ implies the following time evolution of the
Madelung exponents:
$$ \partial _tR = HS - \int_R [exp(R_{xy})\, sin\, S_{xy}\, - \,
S_{xy}]d\nu
(y)$$
$${\partial _tS = - HR\, +\, \int_R [exp(R_{xy})\, cos\, S_{xy}\,
-\, 1\, -\,
R_{xy}]d\nu (y)}\eqno (39)$$
where $H=F(-i\nabla)$.

By employing (38) with respect to $\rho ^{1/2}=exp(R)$, we arrive at:
$${Q:= {{H\rho ^{1/2}}\over {\rho ^{1/2}}} = HR \, -\,
\int_R [exp\, (R_{xy}
)\, -\, 1\, -\, R_{xy}]d\nu (y)}\eqno (40)$$
and hence:
$${\partial _tS  = - Q + \int_R exp(R_{xy})\, [cos (S_{xy})\,
-\, 1]d\nu (y)}
\eqno (41)$$

The same procedure can be  repeated for $\Theta =exp(R+S)$ and $\Theta _*=
exp(R-S)$, where equations (39), (41) imply:
$${\partial _t\Theta  = H\Theta \, +\, \Theta [ - 2Q\, +\,
\int_Rexp(R_{xy})
[-sin\, S_{xy} + cos\, S_{xy} + exp(S_{xy}) - 2]d\nu (y)]}\eqno (42)$$
$${\partial _t\Theta _* = - H\Theta _*\, +\, \Theta _*[2Q \, -\,
\int_R exp(
R_{xy})[sin\, S_{xy}\, +\, cos\, S_{xy} \, +\, exp(-S_{xy})\, -\, 2]
d\nu (y) ]}$$

In contrast to the Gaussian case \cite{garb,blanch},  equations (42)
do not
take the form of a  time adjoint pair, unless some additional restrictions
are imposed on the Madelung exponent $S(x,t)$ (notice that we have restored
 time dependence, skipped before for convenience).
An obvious demand is $S(x+y,t)=S(x,t)$ for all $y, t$, and any fixed $x$.
But then, equations (42) would manifestly refer to the \it stationary \rm
(measure preserving) random dynamics, governed by the pair of equations:
$$\partial _t\Theta = H\Theta - 2Q\Theta $$
$${\partial _t\Theta _* = - H\Theta _* + 2Q\Theta _* }\eqno (43)$$
which are mutually time adjoint. Hence they would fall into the
Schr\"{o}dinger problem framework, with a trivial implication that the
measure preserving process is Markovian. This however cannot be a property
of the "free" dynamics since we need external potentials to secure
stationarity. Let us therefore
make  an essential amelioration by performing the previous  analysis
for the
case $i\partial _t\psi =(H+V)\psi $ with $V=V(x)$. Then, the stationary
system of equations (43) would take the form:
$${\partial _t\Theta =H\Theta - (2Q+V)\Theta }\eqno (44)$$
$$\partial _t\Theta _* = - H\Theta _* + (2Q+V)\Theta _*$$
which upon substituting $S(x,t)=Et$, where $E$ is a constant, yields a
 pseudodifferential  version of the Sturm-Liouville problem:
$${H\rho ^{1/2}(x) - [2{{H\rho ^{1/2}}\over {\rho ^{1/2}}} + V(x) - E]
\rho ^{1/2}(x) = 0}\eqno (45)$$
$$\Downarrow $$
$$V(x)\, -\, E\, =\, -\, {{H\rho ^{1/2}(x)}\over {\rho ^{1/2}(x)}}$$
to be solved (for a chosen value of $E$) with respect to the square
root of
the probability density $\rho (x)$, once the external force potential
$V(x)$ is selected.

This problem has its Gaussian
counterpart in the study of the measure preserving dynamics
\cite{blanch,var}
, and  in the present context it can be solved by invoking those
potentials
for the original pseudodifferential Schr\"{o}dinger equation, for
which the
bound states (i.e., stationary solutions) have granted the existence
status.
The relevant analysis has been carried out in the studies of the
relativistic
stability of matter \cite{herb,carm1,daub,lieb}.  In addition we know
\cite{ichi1,carm1} that in the stationary case, the Feynman-Kac path
integral
generalization to L\'{e}vy semigroup kernels is available.

However, the Markov property cannot automatically be attributed to the
nonstationary dynamics,
as described  by (42). Below, we shall make a careful analysis of the
Cauchy - Schr\"{o}dinger ($H=|\nabla |$) dynamics, to produce a definite
counterexample, for which the unrestricted equations (42) would hold true,
but  the associated random dynamics would be non-Markovian.

\section{The Cauchy-Schr\"{o}dinger dynamics}

\subsection{Construction of an explicit nonstationary solution}

While it is clear that  $exp(-t|\nabla |)$ and $exp(-it|\nabla |)$
have a
common, identity operator limit as $t\downarrow 0$, an analytic
continuation
of the Cauchy kernel by means of (28) gives rise to:
$${k_t(x)={1\over {\pi }} {t\over {x^2+t^2}}\longrightarrow  }\eqno (46)$$
$$k_{is}={1\over 2}[\delta (x-s) + \delta (x+s)] + {1\over {\pi }}
{\cal P}{is
\over {x^2-s^2}}$$
Here, we use the usual notation for the Dirac delta functionals,
and the
new time label $s$ is a remnant of the limiting procedure $t\downarrow 0 $
in $\sigma =t+is$.  The function  denoted by $is/\pi (x^2-s^2)$ comes
from the
inverse Fourier transform of $-{i\over \sqrt {2\pi }}sgn (p) sin (sp)$.
Because of
$${[sgn (p)]^{\vee } = i\sqrt {2\over \pi } {\cal P}({1\over x})}
\eqno (47)$$
where ${\cal P}({1\over x})$ stands for the functional defined in
terms of a
principal value of the integral. Using the notation $\delta _{\pm s}$
for the
Dirac delta functional $\delta (x \mp s)$:
$${[sin (sp)]^{\vee } = i\sqrt {\pi \over 2} (\delta _s - \delta _{-s})}
\eqno (48)$$
we realize that
$${{1\over \pi }{is\over {x^2-s^2}} = {i\over {2\pi }}(\delta _s -
\delta _{-s})*{\cal P}({1\over x})}\eqno (49)$$
is given in terms of the implicit  convolution of two generalized functions.

Let $\psi (x,0)= f(x) := \sqrt {2\over \pi }\, {1\over {1+x^2}}$
be a $L^2$
normed function, which we take as the initial data for the
Cauchy-Schr\"{o}dinger evolution.

With  the unitary kernel $k_{is}$ in hand, we can define the pertinent
evolution in terms of a convolution  $\psi (x,s):=f*k_{is}=
{i\over \sqrt {2\pi }}
f*(\delta _s-\delta _{-s})*{\cal P}({1\over x})$. Let us consider
$${{i\over {2\pi }}\, f*\delta _s*{\cal P}({1\over x})\, =\,
{i\over {2\pi }}
\, {\cal P} \int_{-\infty }^{+\infty }{f(x-s-y)\over y}\, dy\, = }
\eqno (50)$$
$${i\over {2\pi }}lim_{\epsilon \downarrow 0}
[\int_{-\infty }^{-\epsilon }
{f(x-s-y)\over y} dy \, +\, \int_{\epsilon }^{\infty } {f(x-s-y)\over y}
dy]$$

Because of:
$${\int [{1\over {1+(x-s-y)^2}}]\, {dy\over y}\, =\, A\, ln\, [{|y|
\over \sqrt {1+(x-s-y)^2}}]\, -\, A\, (x-s)\, arctan\, (x-s-y)}
\eqno (51)$$
where
$${A\, =\, {1\over {1+(x-s)^2}}\, ,}\eqno (52)$$
the definite integrals in (50) read:
$${\int_{-\infty }^{-\epsilon }\, {{f(x-s-y)}\over y}\, dy \, =\,
\sqrt {2\over {\pi }}\, A\, [ln\,
{\epsilon \over {\sqrt {1+(x-s+\epsilon )^2}}}\,
-\,}$$
$${(x-s)\, arctan\, (x-s+\epsilon ) \, +\, {\pi \over 2}(x-s)]}
\eqno (53)$$
$$\int_{\epsilon }^{\infty }\,  {{f(x-s-y)}\over y}\, dy\, =\,
\sqrt {{2\over \pi }}\, A\, [-\, ln\,
{\epsilon \over \sqrt {1+(x-s-\epsilon )^2}}\, +\,$$
$${\pi \over 2}(x-s)\, +\, (x-s)\, arctan\, (x-s-\epsilon )]$$
and therefore:
$${{\cal P} \int_{-\infty }^{+\infty }\, {{f(x-s-y)}\over y}\, dy \, =\,
{\sqrt {2\pi }(x-s)\over {1+(x-s)^2}}}\eqno (54)$$
By proceeding analogously (with $\delta _{-s}$ replacing $\delta _s$ in
(50)) we find:
$${{\cal P} \int_{-\infty }^{+\infty } \, {{f(x+s-y)}\over y}\, dy\,
=\, {\sqrt {2\pi }(x+s)\over {1+(x+s)^2}}}\eqno (55)$$
All that finally implies (remembering that $f(x)={\sqrt {2\over \pi }
}/(1+x^2)= \psi (x,0)$):
$${\psi (x,s) = [exp(-isH)\, f](x)\, =\, {1\over 2}\, [f(x+s)\, +\, f
(x-s)]\, +\, }$$
$${{i\over 2}\, [(x-s)f(x-s)\, -\, (x+s)f(x+s)]}\eqno (56)$$
with an interesting formula for the time developement of the
probability density:
$${\rho (x,s) :=\, |\psi (x,s)|^2\, =\, (1+s^2)\, \sqrt {\rho _0(x+s)
\rho _0(x-s)}}\eqno (57)$$
$$\rho _0(x)\, =\, |\psi (x,0)|^2 = [f(x)]^2 \,=\, {2\over {\pi }}
{1\over (1+x^2)^2}$$
Notice that by a direct evaluation, we can check the normalization
identity $\int_{-\infty }^{+\infty } \rho (x,s)\\ dx=1$.

Now, we can address the problem of whether the stochastic process,
implying the propagation (57) of the probability density, is a Markov
process.

\subsection{The nonstationary Cauchy-Schr\"{o}dinger stochastic
process is not Markov}

\subsubsection{Candidates for the transition probability  density}

We are interested in representing the time evolution of $\rho (x,t)$,
(57), in the integral (probabilistic transport rule-looking) form:
$${\rho (x,t)\, =\, \int_R \, p(y,0,x,t)\, \rho _0(y)\, dy\, = \,
\int_R \, p(y,s,x,t)\, \rho (y,s) \, dy}\eqno (58)$$
without bothering at the moment whether we can assign to $p(y,0,x,t)$
or $p(y,s,x,t),\, \\ s<t$, any true meaning of the transition probability
density of a stochastic process.

Since $\rho (x,t)$ \it is \rm a probability density, we can evaluate
its characteristic function (Fourier transform \cite{luk}):
$\phi (p,t):= \sqrt {2\pi }\hat{\rho }(p,t)$.  Of course, $\phi (p,0)=
\hat{f}(p)= \int_R [1/\pi (1+x^2)]exp (ipx)\, dx=exp (-|p|)$ is a
characteristic function  as well.

By observing that ($\psi (x,0):=\psi _0(x)$)
$${{1\over 2}\, [\psi _0(x+t) + \psi _0(x-t)]\, =\, \sqrt {2\over \pi }\,
{{(1+t^2)+x^2}\over  {[1+(x+t)^2][1+(x-t)^2]}}}\eqno (59)$$
while
$${{1\over 2}[(x-t)\psi _0(x-t) - (x+t)\psi _0(x+t)]\, =\, t\,
\sqrt {2\over \pi}\, {{-(1+t^2)+x^2}\over {[1+(x+t)^2][1+(x-t)^2]}}}
\eqno (60)$$
we arrive at  (cf. (56) for the definition of $\psi (x,t)$)
$${Re\, \psi (x,t)\, - \, {1\over t}\, Im\, \psi (x,t)\, =\, 2
\sqrt {2\over \pi }\, {{(1+t)^2}\over {[1+(x+t)^2][1+(x-t)^2]}}\, =\,
\sqrt {2\pi }\,
\rho (x,t)}\eqno (61)$$
In view of (61), we have
$${\hat{\rho }(p,t)\, =\, {1\over \sqrt {2\pi }}\int_R exp(ipx)\,
\rho (x,t)\, dx\, =\, }$$
$${{1\over \sqrt {2\pi }}\int_R exp\, (ipx)\, [Re\, \psi (x,t)\, -\,
{1\over t}\, Im\, \psi (x,t)]\, dx}\eqno (62)$$
where
$${\psi (x,t)\, =\, {1\over \sqrt {2\pi }}\int_R exp(\pm iqx - |q| - i
|q|t)\, dq}\eqno (63)$$
So that
$${{1\over \sqrt {2\pi }}\phi (p,t)\, =\, \hat{\rho }(p,t)\, =\,
{1\over \sqrt {2\pi }}\, exp\, (-|p|)\, [cos\,(t|p|)\, +\,
{1\over t}\, sin\,
(t|p|)]}\eqno (64)$$

Finally, in view  of $\hat{\rho }_0(p)=(1/\sqrt {2\pi })(1+|p|)
exp(-|p|)$,
we find  ($t\geq 0$):
$${\hat{\rho }(p,t)\, =\, \hat{\rho }_0(p)\, {{cos\, (t|p|)\, +\,
{1\over t}\, sin\, (t|p|)}\over {1\, +\, |p|}}}\eqno (65)$$
and ($0<s<t$):
$${\hat{\rho }(p,t)\, =\, \hat{\rho }(p,s)\, {{cos (t|p|)\, +\,
{1\over t}
sin (t|p|)}\over {cos (s|p|)\, +\, {1\over s} sin (s|p|)}}}\eqno (66)$$

Ignoring the issues of existence and
positive--definiteness of Fourier transformed integrands, we can
proceed in the standard way:
$${\rho (x,t)\, =\, {1\over \sqrt {2\pi }}\int_R exp (-ipx)\, \hat{\rho }
(p,t)\, dp\, =\, \int_R p(y,s,x,t)\rho (y,s) dy\, = }$$
$${\int_R [{1\over {2\pi }} \int_R  exp[ip(y-x)] {{cos (t|p|)+
{1\over t}sin
(t|p|)}\over {cos (s|p|) +{1\over s} sin (s|p|)}}\, dp\, ]\, \rho (y,s)\,
dy}\eqno (67)$$
where the formal, homogeneous in space "transition probability density"
$p(y,s,x,t),\, \\ s<t$, trivially satisfies the formal
Chapman-Kolmogorov identity: $p(y,s,x,t)\\ =\int_Rp(y,s,z,u)p(u,z,x,t)dz\,
,\, s<u<t$.

To go beyond  formal arguments:

(i)  We need to prove that the function $p(y,0,x,t),\, t\geq 0$ is a well
defined transition probability density of the stochastic process
transporting the initial (time $0$) density into the terminal
(time $t\geq 0$) one.

(ii) We need to demonstrate that the would-be transition density
$p(y,s,x,t),\, s<t $,  is a well defined probability measure, and actually
we shall
prove that it is \it not \rm , which excludes  the Markov property for the
stochastic process under consideration in agreement with our previous
conclusions of Section 2.

\subsubsection{Existence of  the  probabilistic transport from time $0$ to
time $t\geq 0$}

 In agreement with (65)-(67), we can introduce an integral kernel
 $p(y,0,x,t)$  effecting the transport of $\rho (y,0)$ into $\rho (x,t)\,
 ,\, t\geq 0$:
$${p(y,0,x,t)\, = \, {1\over {2\pi }}\int_R exp[ip(y-x)]\, {{cos(t|p|)\,
+\, {1\over t}\, sin(t|p|)}\over {1\, +\, |p|}} dp} \eqno (68)$$
At the moment, its status as the transition probability density of a
stochastic process is not settled. We must know whether its Fourier
transformed
integrand  is positive-definite function and satisfies a normalization
identity
$\int_R p(y,0,x,t)\, dx \, =\, 1$ for all times $t\geq 0$.

In view of the homogeneity in space we observe that  $p(y,0,x,t)=
p(0,0,x-y,t)$  and so we  pass to  the notation
$${\sqrt {2\pi }\, p(x,t) = [{{cos(tp)}\over {1+|p|}}\, +\, {{|p|}
\over {1+|p|}}{{sin(tp)}\over {tp}}]^{\vee }}\eqno (69)$$
The function
$${g(x):=\, {1\over \sqrt  {2\pi }}({1\over {1\, +\, |p|}})^{\vee }(x)\,
=\, {1\over {2\pi }}\int_R exp (-ipx)\, {dp\over {1+|p|}}}\eqno (70)$$
will play a distinguished r\^{o}le in what follows. Indeed,  because
$${[cos(tp)]^{\vee } = \sqrt {\pi \over 2}(\delta _t\, +\, \delta _{-t})}
\eqno (71)$$
$$({{sin(tp)}\over {tp}})^{\vee } \, =\, {1\over t}\,
\sqrt {\pi \over 2}\,
\chi _{[-t,t]}$$
where $\chi _{[-t,t]} = 1 $ if $|x|\leq t$, and $=0$ if $|x|>t$, we can
rewrite $p(x,t)$ as follows:
$${p(x,t)\, =\, {1\over 2}[g(x+t)\, +\, g(x-t)]\, +\, }\eqno (72)$$
$${1\over 2t}\, \chi _{[-t,t]}(x)\, -\,
{1\over {2t}}(g*\chi _{[-t,t]})(x)
$$
In view of (72), properties of $p(x,t)$  are completely determined by
those of $g(x)$.

Eq.(70) suggests that $g(x) $ itself might be a probability density.
For this to hold true, $1/(1+|p|)$ must be a characteristic function,
and we are in the framework covered  by the classic Bochner Theorem
\cite{bre,boch,loev}.
Our integrand $1/(1+|p|)$ is continuous on $R$ and equals $1$
when $|p|=0$.
It is well known \cite{bre} that such a function is a characteristic
function if and only if  it is positive definite.
To be a positive definite
 function $h(p)$ must satisfy the inequality
$${\Sigma _{i,j,=1}^nh(p_i\, -\, p_j)\, \lambda _i
\overline  {\lambda }_j\,
\geq \, 0 }\eqno (73)$$
for any finite sequence of complex numbers $\lambda _1,\lambda _2,...,
\lambda _n$, any sequence of points $p_1,p_2,...,p_n$ in $R$ , and any
$n=1,2,...$.

The identity (73) is trivially satisfied by $h(p)=1/(1+|p|)$, and as a
consequence it is a  characteristic function of the probability
density $g(x)$,
(70). Then $\int_R g(x)dx=1$ follows.

The function $p=p(x,t)$ is real and even in $x$, which allows us to
consider
a nonnegative semiaxis $x\geq 0$ for the moment. Let $0\leq x<t$.
Because $\int_R g(x)dx=1$, we find that
$${\int _{x-t}^{x+t} g(y)dy \, <\, 1\Longrightarrow   {1\over {2t}}\, -\,
{1\over {2t}} (g*\chi _{[-t,t]})(x)\, >\, 0 }\eqno (74)$$
which implies that
$${p(x,t)\, =\, {1\over 2}[g(x+t) + g(x-t)]\, +\, {1\over {2t}} \, -\,
{1\over {2t}} (g*\chi _{[-t,t]})(x) \, > \, 0}\eqno (75)$$
If $x>t$ it suffices to consider
$${1\over {2t}}\int _{x-t}^{x+t} g(y) dy\, <\, {{g(x+t)+g(x-t)}\over 2}
\Longrightarrow $$
$${{1\over 2}\, [g(x+t)\, + g(x-t)] \, -\, {1\over {2t}}
(g*\chi _{[-t,t]}(x)\, >\, 0\, \Longrightarrow
 p(x,t)\, > 0}\eqno (76)$$

Hence, $p(x,t)$, (72) is  strictly positive on R, and moreover it has a
unit normalization. This implies that $p(0,0,x-y,t)$ is a well defined
probability density.

Let us add that $lim_{|x|\rightarrow \infty }\, p(x,t) = 0$ and
$lim_{t\downarrow 0}\, p(x,t) = \delta (x)$. At the points $x=\pm t$,
$p(x,t)$
develops singularities in view of $lim_{x\downarrow 0}\,  g(x)= \infty $.
The location of the singularities is quite significant, since they make a
clear distiction between the
random propagation regimes with jumps of size less than $t$, and those of
size greater than $t$, for each terminal time instant of the evolution of
$\rho (x,t)$.

Notice also that as a suitable  probability density $p(x,t)$ leads to
several
finite moments: $\int_Rxp(x,t)dx = 0\, , \, \int_R x^2p(x,t)dx = t^2$.
On the other hand, the long--tailed Cauchy transition density (34) has no
finite moments at all.

\subsubsection{Violation of Markov property}

In view of our previous discussion we have specified consistent
distribution functions for the process, which in principle permits its
construction ( \cite{bre}, chap.15.). For the Markov process what
needs to be specified \cite{bre,loev,gih} are the transition
probabilities for all intermediate times of the considered evolution
and the initial distribution (which is given in our case). We shall
investigate the existence of the \it consistent \rm transition
probability densities for the process in question.

If the
Cauchy-Schr\"{o}dinger process is to be Markov, the integral kernel:
$${p(y,s,x,t)\, =\, {1\over {2\pi }} \, \int_R\, exp[ip(y-x)]\,
{{cos(t|p|)
\, +\, {1\over t}sin(t|p|)}\over {cos(s|p|)\, +\, {1\over s}sin(s|p|)}}\,
dp}\eqno (77)$$
must be the Fourier inversion of a characteristic function.  A necessary
condition for
$${h(p,s,t)= {{cos(t|p|)\, +\, {1\over t}sin(t|p|}\over {cos(s|p|)\, +
{1\over s}sin(s|p|)}}}\eqno (78)$$
to be a characteristic function is the positive-definiteness of
$h(p,s,t)$ as a function of $p\in R$  for all intermediate time instants
$0<s<t$.

We shall prove that for each terminal time instant $t$ we can single out
earlier time instants $s$, such that $h(p,s,t)$ is \it not \rm a
positive-definite function.
As a consequence, the intermediate propagation cannot be Markov.

To this end, let us notice that for a fixed  $t$, and $s>0$ we can
rearrange the denominator in  (78):
$${cos(s|p|)\, +\, {1\over s}sin(s|p|)\, =\, {1\over {cos\, \alpha _s}}\,
cos(s|p|\, -\, \alpha _s)}\eqno (79)$$
$$tan\, \alpha _s:=\, {1\over s}$$
Notice that
$${|p|\, =\, {1\over s} (\alpha _s\, +\,  {{2N+1}\over 2}\, \pi )
\Longrightarrow  cos(s|p|\, -\, \alpha _s)\, =\, 0}\eqno (80)$$
for all integer $N$.

Our $h(p,s,t)$ should satisfy the condition (73). Let us choose the
simplest
case of $n=2$ in this formula, and consider
$\Sigma _{i,j=1}^2 h(p_i\, -\,
p_j)\lambda _i\overline {\lambda }_j \geq 0 $. Because of  (78),
the two by
two matrix $h_{ij}:=h(p_i-p_j)\, ,\, i,j,=1,2$ has matrix elements
$h_{11}=1=h_{22}$ and
$h_{12}=h_{21}=M$ with:
$${M\, =\, cos(\alpha _s)\, {{cos(t|p_1-p_2|)\, +\, {1\over t}
sin(t|p_1-p_2|)}\over {cos(s|p_1-p_2|\, -\, \alpha _s)}}}\eqno (81)$$
To have $det[h_{ij}] \geq 0$ we need $|M|\leq 1$. This condition
can always be
violated by choosing any pair  $p_1,p_2$ for which, at a given time
$s$, the numerical value of $|p_1-p_2|=|p|$ is close to any
of those introduced in the formula (80).

Therefore, the  considered stochastic process (with the transition
mechanism (67)) is not Markov, as anticipated on the basis of arguments
of Section 2.

\section{Meaning of the pseudodifferential stochastic propagation:
an insight into jump features of the process}

\subsection{Fokker-Planck equations}

The probability density $\rho (x,t)$ (respectively $\overline {\rho }
(x,t)$)
was a fundamental entity in all our previous considerations: either (i)
providing the input--output statistics data for the
Schr\"{o}dinger--random
dynamics reconstruction problem, or  (ii) providing  the time evolution of
the probability measure for the whole time interval of interest, so
that the
transition probability densities could be sought for.

In the Gaussian case of Remark 1 we dealt with the temporal evolution of
the probability density given in its traditional Fokker--Planck form
appropriate for Markov diffusion processes. In connection with the
pseudodifferential ("free noise") dynamics, equations (24),(25) are an
obvious  extension of the previous notion to a class of jump processes. We
shall extend the usage of the name  Fokker--Planck equation to any first
order in time  differential equation determining the space-time properties
of $\rho (x,t)$ or $\overline {\rho }(x,t)$.

Let us investigate the time developement of $\overline {\rho }(x,t)=
\theta (x,t)\theta _*(x,t)$, where  $\theta (x,t),\, \\ \theta _*(x,t)$
come
out as solutions of the temporally adjoint  pair of equations of the form
$${\partial _t\theta = H\theta  - V\theta }\eqno (82)$$
$$\partial _t\theta _* = -H\theta _* + V\theta _*$$
with  the initial/terminal data $f(x),\, g(y)$ of the Schr\"{o}dinger
problem (19),(20) and a Feynman-Kac potential $V$. Then, in view of (37)
and
$\theta =exp(R+S)$, $\theta _*=exp(R-S)$, we get an evolution equation
for the probability density:
$${\partial _t\overline {\rho }(x,t) = \theta _*(x,t)\, (H\theta )(x,t)\,
-\, \theta (x,t)\, (H\theta _*)(x,t)\, =}\eqno (83)$$
$$\int _R [-\theta _*(x,t)\theta (x+y,t) + \theta (x,t)\theta _*(x+y,t) +
2\overline {\rho }(x,t) \nabla S(x,t) {y\over {1+y^2}}] d\nu (y)$$
Following the traditional recipes when dealing with L\'{e}vy measures
\cite{bre}, let us consider an open neighborhood of the origin
$|\epsilon |\ll 1$.
Instead of integrating over all possible jump sizes, let us integrate over
jumps of size $|y|>\epsilon >0)$. The removal of this lower bound as
$ \epsilon \rightarrow 0$ will eventually amount to evaluating the
principal
value of the integral. In case  $\epsilon >0$, we can safely remove the
compensating term  including $y/(1+y^2)$ from the integral, and restrict
considerations to the contribution from the first two terms only.

Our purpose is to establish a connection between (83) and the conventional
theory of jump stochastic processes, as developed in \cite{gih}.
Integrating
over a Borel set $A\subset R\, ,\, x\in A$ we get:
$$\int _A\, dx \, \int _{|y|>\epsilon }[-\theta _*(x,t)\theta (x+y,t) \,
+ \theta (x,t)\theta _*(x+y,t)]d\nu (y)\, = $$
$${\int_R dx\, \int_{|y|>\epsilon }  \, \chi _A(x)\, [-\,
\overline {\rho }
(x,t)\, {{\theta (x+y,t)}\over {\theta (x,t)}}\, +\,
\overline {\rho }(x+y,t)\, {{\theta (x)}\over {\theta (x+y)}}]\,
d\nu (y)\, =\, }\eqno (84)$$
$$\int_R\, dx\, \overline {\rho }(x,t) \, \int_{|y|>\epsilon }\,
{{\theta (x+y,t)}\over {\theta (x,t)}}[\chi _A(x+y)\, -\, \chi _A(x)]\,
d\nu (y)$$
where we interchanged the order of integrations, and made appropriate
adjustments of integration variables ($x\rightarrow x-y$ and
$y\rightarrow -y$), while exploiting the property $d\nu (-y)=-d\nu (y)$ of
measures (35),(36); $\chi _A(x)$ is an indicator function of the
Borel set
$A\subset R$, equal to $1$ when $x\in A$ and $0$ otherwise.

In the present case, (82), we deal with a Markov process with transition
probability densities given for arbitrary time instants:
$\overline {\rho }(x,t)=\int_R
 p(y,s,x,t)\overline {\rho }(y,s)dy,\, s<t$. By invoking the standard
 wisdom
 about jump Markov processes \cite{gih}, and exploiting
 $lim_{t\downarrow s}p(y,s,A,t)=\chi _A(y)$, for any Borel set
 $A\subset R$
 away from $(-\epsilon ,+\epsilon )$, we can define the jump process
 running
 with jumps of size  $|y|>\epsilon >0$. It should be viewed as an
 approximation of the original stochastic process governed by (83),
 with the
 initial data $\overline {\rho }(x,0)$ common for both:
 $${\partial _t\overline {\rho }_{\epsilon }(A,t)\, =\, \int_R q(x,t,A)\,
 \overline {\rho }_{\epsilon }(x,t) \, dx \, +\, \langle v\rangle _A(t)
 \int_{|y|>\epsilon } {y\over {1+y^2}} d\nu (y)}\eqno (85)$$
 where
 $${q(x,t,A):= lim_{u\downarrow t}{1\over {u-t}}\, [p(x,t,A,u)\, -\,
 \chi _A(x)] =}$$
 $${\int_{|y|>\epsilon }{{\theta (x+y,t)}\over {\theta (x,t)}}
 [\chi _A(x+y)-
 \chi _A(x)]d\nu _(y),}\eqno (86)$$
 $$\langle v\rangle _A(t) := \int_A \overline {\rho }(x,t)\,
 [2\nabla S(x,t)]\, dx$$
 Here $q(x,t,A)\geq 0$ for all $x$ which are \it not \rm in $A$, in
 agreement with \cite{gih}.
We have also introduced a  pseudodifferential counterpart of the current
velocity field $v(x,t)=2\nabla S(x,t)$, previously attributed to
diffusion
processes (cf. Remark 1), where the probability conservation law
(a continuity equation in another lore) $\partial _t\rho =
-\nabla (v\rho )$
plays the r\^{o}le of the Fokker--Planck equation.

Notice, that  in the particular case of  $\theta (x,t)\equiv 1$ for all
$x,t$, and $V=0$, Eq.(82) reduces to the "free noise" situation covered by
the Fokker-Planck equations (24),(25). Then,
$q(t,x,A)= \int_{|y|>\epsilon } [\chi _A(x+y)\, -\, \chi _A(x)]d\nu _(y)$,
while $R=-S\, ,\, \overline {\rho }=exp(2R)=\theta _*$ implies
$\langle v\rangle _A(t)=- \overline {\rho }(x,t)|_a^b$ where
$[a,b]:=A\subset R$.

Now, let us address the Fokker--Planck  equation for the
pseudodifferential--
Schr\"{o}\-dinger dynamics case, which we consider in the form analogous to
(82); see also (1) for comparison:
$${i\partial _t\psi = H\psi + V\psi}\eqno (87)$$
$$i\partial _t\overline {\psi } = -H\overline {\psi } -
V\overline {\psi }$$
We re-emphasize  that to define the probability density
$\rho (x,t)=|\psi (x,t)|^2$ one actually employs solutions of  the
time adjoint pair of Schr\"{o}dinger equations.

In view of (87), the pseudodifferential continuity equation follows:
$${\partial _t\rho (x,t) = -i[\overline {\psi }(x,t) (H\psi )(x,t)\,
-\, \psi (x,t) (H\overline {\psi })(x,t)]\, =}\eqno (88)$$
$$-i\int_R [-\overline {\psi }(x,t)\psi (x+y,t) + \psi (x,t)
\overline {\psi }
(x+y,t) + 2i\rho (x,t)\nabla S(x,t){y\over {1+y^2}}]d\nu (y)$$
Our next  step is a repetition of the procedures behind (84), which
implies:
$${\partial _t\rho (x,t) = \int_R [2{\cal I}[\psi (x,t)\overline {\psi }
(x+y,t)]+ 2\rho (x,t)\nabla S(x,t) {y\over {1+y^2}}]d\nu (y)}$$
$$\Downarrow $$
$${\int_A\, dx\, \int_{|y|>\epsilon } 2{\cal I}[\psi (x,t)
\overline {\psi }
(x+y,t)]=}$$
$${\int_R\, dx\, \int_{|y|>\epsilon } \chi _A(x) 2\rho ^{1/2}(x,t)
\rho ^{1/2}
(x+y,t)\, sin[S(x,t)-S(x+y,t)] d\nu (y)\, =\, }\eqno (89)$$
$$\int_R\, \rho (x,t) dx\, \int_{|y|>\epsilon } {{\rho ^{1/2}(x+y)}
\over {\rho ^{1/2}(x)}} sin[S(x+y,t)-S(x,t)] [\chi _A(x+y)-\chi _A(x)]
d\nu(y)$$
where ${\cal I} [f(x,t)]$ stands for an imaginary part of a complex
function $f(x,t)$. So, a counterpart of (85) reads:
$${\partial _t\rho _{\epsilon }(A,t)\, =\, \int_R q(x,t,A)
\rho _{\epsilon }
(x,t) dx + \langle v\rangle _A(t)\int_{|y|>\epsilon }
{y\over {1+y^2}}d\nu (y)}\eqno (90)$$
where, however
$${q(x,t,A) := \int_{|y|>\epsilon } {\cal I} [{{\psi (x+y,t)}
\over {\psi (x,t)}}]\, [\chi _A(x+y)-\chi _A(x)]d\nu (y)}\eqno (91)$$
no longer can be derived from transition probability densities of the
process,
as in the previous discussion (86), because in general our process  is
\it not \rm Markovian. At least in the case of  nonstationary dynamics,
the
only transition probability density which is at our disposal connects an
initial instant of the evolution with any later one.
In fact, we might even not  be sure that  $q(x,t,A)$ is a well defined
probabilistic object, because of the presence of $sin[S(x+y,t)-S(x,t)]$ in
the integrand. At this point an observation  of \cite{ang} helps. Namely,
in view of the identity:
$${\int_Rdx\int_{|y|>\epsilon } |\psi (x+y,t)\psi (x,t)|[\chi _A(x+y)-
\chi _A(x)] d\nu (y)\, =\, 0}\eqno (92)$$
valid for Borel sets $A\subset R$, which are away from $(-\epsilon ,
+\epsilon )$, we can always pass from (89) to the rearranged form of (91):
$${q(x,t,A)=\int_{|y|>\epsilon } [\, |{{\psi (x+y,t)}\over {\psi (x,t)}}|
+ {\cal I} [{{\psi (x+y,t)}\over {\psi (x,t)}}]\, ]\, [\chi _A(x+y)-
\chi _A(x)]\, d\nu _(y)}\eqno (93)$$
implying that $q(x,t,A)$ is positive for all $x$ which are \it not \rm in
$A$, as should be the case \cite{gih}.

\subsection{The jump processes toolbox}

To have a better insight into stochastic jump processes associated with the
Fokker--Planck evolutions (83), (85) and (88), (90) respectively, some
further knowledge of the general theory is necessary. We shall try to
minimize the level of sophistication by invoking arguments based on the
exploitation of the standard Poisson process.

A random variable  $X$ taking discrete values $0,y,2y,3y,...$, with $y>0$
is said to have Poisson distribution ${\cal P}(\lambda ),\, \lambda
\geq 0$
with jump size $y$, if  the probability of $X=ky$ is given by
$P(X=ky)= {{\lambda ^k}\over {k!}}exp(-\lambda )$. The characteristic
function of ${\cal P}(\lambda )$ reads:
$${E[exp(ipX)] = exp[\lambda (e^{ipy} - 1)] = \Sigma _0^{\infty }
e^{-\lambda }\, {{\lambda ^k}\over {k!}}e^{ik(py)}\, =\,
\Sigma _0^{\infty } e^{ik(py)}\, P(X=ky)}\eqno (94)$$
and  its first moment equals $E[X] = \lambda $. Notice that $P(X=0)=
exp(-\lambda )$, hence the numerical value of $\lambda \geq 0$ tells us
what is the probability of a jump \it not \rm to occur at all for a
given Poisson process.
For the Poisson random variable with values $b+ky, k=0,1,...$we would get
$${E[exp(ipX)]= exp[ibp +\lambda (e^{ipy} - 1)]\, .}\eqno (95)$$
If we consider  $n$ independent random  variables $X_j, 1\leq j\leq n$ such
that $X_j$ has  Poisson distribution ${\cal P}(\lambda _j)$ with jump
size $y_j$, then a new process $X$ can be introduced with the distribution
of $X_1+...+X_n$ so that its characteristic function reads
$${E[exp(ipX)]= exp[\Sigma _{j=1}^n \lambda _j(e^{ipy_j} - 1)]}
\eqno (96)$$
The exponent in (96) would include an additional term
$ip\Sigma _1^n b_j$ if
nonrandom shifts of each jump $ky_j$ by $b_j$ would be allowed.

We can admit not only jumps of fixed magnitudes $y_1,...,y_n$ but also
jumps covering an arbitrary range in $R_+$. Let the distribution function
of the magnitude of the jump be $P(x<y)=\mu (y)$.  A possible
generalization of (96) to this case is
$${E[exp(ipX)]= exp[\int_{R_+} (e^{ipy} - 1) d\mu (y)]}\eqno (97)$$
assuming that the integral in the exponent exists.  Notice that (96) is
recovered, if we set
$${d\mu (y)=\Sigma_{j=1}^n \lambda _j\delta (y-y_j)\, dy}\eqno (98)$$

The convergence of the exponent in (97) may be jeopardized in cases when
jumps of very small amplitude are allowed to occur very often, while we
take for granted that jumps of very large size  seldom happen.
On the other hand \cite{bre}, for  any Borel set $A\subset R $
bounded away
from the origin, the process $X_A$ of jumps bounded by $A$, has the
characteristic exponent $\int_A (e^{ipy}-1)d\mu (y)$, and the expected
number $E_A[X]$ of jumps of size $A$ is equal to $\mu (A)$. We can
say that
the processes of jumps of different sizes proceed independently of one
another, and the jump process of jumps of size $[y,y+\triangle y), \,
\triangle y\ll 1$ contributes a Poisson component with exponent function
approximately equal to $(e^{ipy}-1)\mu ([y,y+\triangle y))$. At the
moment the
processes have only upward jumps, hence their sample paths are
nondecreasing.

For a process with the characteristic exponent $-F(p)$, (23), we can
consider
its restriction to upward jumps  of size $y>\epsilon >0$
$${\phi _{\epsilon }^+(p) = \int_{y>\epsilon } [e^{ipy}-1- {{ipy}
\over {1+y^2}}]d\nu (y) = \int_{y>\epsilon }[e^{ipy} - 1 ]d\nu (y)\, -\,
ipb _{\epsilon }^+}\eqno (99)$$
$$b_{\epsilon }^+ = \int_{y>\epsilon } {y\over {1+y^2}}d\nu (y)$$
Clearly,  we deal here with a process of the type considered before, and
might try to isolate contributions from jumps  of size
$[y,y+\triangle y)$ by
considering a coarse-graining of a Borel set $A$ of interest.
A formal substitution of (98) in (99), with $d\mu $ replacing $d\nu $,
gives rise to
$${E[exp(ipX)]= exp[\Sigma _{j=1}^n [\lambda _j(e^{ipy_j}-1)
- ip\, {{\lambda _jy_j}\over {1+y^2_j}}]}\eqno (100)$$
to be compared with (95).

Further specializing the problem to relativistic Hamiltonians, we notice
that the corresponding L\'{e}vy measures $\nu _0(y)$ and $\nu_m(y)$,
(35),(36)
are even  under space reflections, hence $d\nu (-y)=-d\nu (y)$ in these
cases. Consequently, we can easily extend our discussion to jumps of
all sorts  in R, i.e. $y$ can take values in both $R_+$ and $R_-$,
with the only
restriction to be observed that $|y|>\epsilon >0$.
Notice that we shall deal with two processes, which run separately  with
either positive or negative jumps, and there is no common jump point for
them. This fact means that they are independent components of the more
general process defined by:
$${\phi _{\epsilon }(p) = \int_{|y|>\epsilon } [e^{ipy} - 1]d\nu (y)
- ipb_{\epsilon }}\eqno(101)$$
where in the case of the L\'{e}vy measures given in (35),(36) the
deterministic term identically vanishes in view of
$${b_{\epsilon }= b_{\epsilon }^+\, +\, b_{\epsilon }^- =
\int_{y>\epsilon }{y\over {1+y^2}}d\nu (y) \, +\, \int_{y<-\epsilon }
{y\over {1+y^2}}d\nu (y) \equiv 0}\eqno (102)$$

All our steps (94)-(101) involved the fact that we deal with infinitely
divisible probability laws. One additional  important property about
them
is that \cite{bre} if $exp\phi (p)$ is a characteristic function of a
given probability distribution, then
$[exp\phi (p)]^t \\ =exp[t\phi (p)],\, t>0$
is likewise a characteristic function of an infinitely divisible
probability law again.
This feature readily extends our discussion to time-dependent stochastic
processes (time homogeneous with independent increments, associated by us
with the "free noise").
Obviously, for such processes $E[exp(ipX(t))]=exp[t\phi (p)]$ while
$E_A[X(t)]=t\nu (A)$,  and our previous arguments retain their validity
with respect to
$${E[exp(ipX(t))]_{\epsilon } = exp[t\phi _{\epsilon }(p)] =
exp[t\, \int_{|y|>\epsilon }(e^{ipy}-1)d\nu (y)]}\eqno (103)$$
$$\partial _t\overline {\rho }(x,t)=-(H\overline {\rho })(x,t)
\Longrightarrow
\partial _t\overline {\rho }_{\epsilon }(A,t)=\int_R dx \,
[\int_{|y|>\epsilon }
 [\chi _A(x+y)-\chi _A(x)]d\nu (y)]\, \overline {\rho }_{\epsilon }(x,t)$$

{\bf Remark 4}: In fact, (102) means  that  the Fokker--Planck equations
(85),(90), if specialized to  L\'{e}vy measures (35),(36), involve
exclusively the integral term on their right-hand-side:
$${\partial _t\overline {\rho }_{\epsilon }(A,t) = \int_R \overline {q}
(x,t,A)\overline {\rho }_{\epsilon }(x,t)dx}\eqno (104)$$
$$\partial _t\rho _{\epsilon }(A,t) = \int_R q(x,t,A)\rho _{\epsilon }
(x,t)dx$$
where an overbar distinguishes between probabilistic quantities
characterising
different families of stochastic jump processes (86) and (91)
respectively.  Let us emphasize that the simplification (104) occurs
only in the $|y|>\epsilon>0$ jumping size regime. The real r\^{o}le
of two terms in (102)
is to compensate the divergent contributions from the L\'{e}vy measure
when the principal value integral $\epsilon \rightarrow 0$ limit is
considered; then the \it standard \rm jump  process theory (104)
does not apply.
Anyway, those two terms  are irrelevant for any $\epsilon >0$,
irrespectively
of  how small $\epsilon $ is.

One might expect that  an infinitesimal
(jump size)  surrounding of the origin gives a dominant jump contribution
to the process.
However, generally it is not the case: explicit solutions (34),(36) and
(57),
(72) indicate that for times $t>0$ the leading contribution does not
necessarily come from jumps of infinitesimal sizes.

\section{Relativistic wave equations and associated sto\-chastic processes}

We  mentioned before that solutions of our
pseudodifferential--Schr\"{o}dinger
equations solve the relativistic wave (or matter--wave in the Klein-Gordon
case) equations as well; see (29)-(32).
Since each particular solution has an undoubted probabilistic significance,
we can re-analyze the old-fashioned problem  \cite{schwe,hol} of
a"single-particle interpretation" for free Klein-Gordon solutions and
analyze the same problem for the D'Alembert equation solutions, from a
novel perspective; see also \cite{ang}.
As well, we can benefit from  relativistic covariance properties of wave
equations to understand how the pseudodifferential--Schr\"{o}dinger
stochastic processes comply with the principles of special relativity.

To begin with, let us consider the Klein-Gordon equation
for a particle of mass $m>0$:
$${ (\Box + m^2)\phi (\vec{x},t) = 0}\eqno (105)$$
The spacetime metric signature is $diag(g_{\mu \nu })
=(1,-1,-1,-1,)$, and the system of units is $\hbar =c=1$.
In view of the polar (Madelung) decomposition of the complex wavefunction,
$\phi (\vec{x},t)=exp[R(\vec{x},t)+iS(\vec{x},t)]$, we can split (105)
into two real equations:
$${(\partial _{\mu }S)(\partial ^{\mu }S) = m^2 +
{{\Box \rho ^{1/2}}\over {\rho ^{1/2}}}}\eqno (106)$$
$$\partial _{\mu }j^{\mu } = 0 $$
$$ j^{\mu }:={1\over {2i}}[\overline {\phi }\partial ^{\mu }\phi  -
\phi \partial ^{\mu }\overline {\phi }]  = -{\rho }
(\partial ^{\mu }S)$$
where $\rho (\vec{x},t) = |\phi (\vec{x},t)|^2=exp[2R(\vec{x},t)]$.

We can handle the  $m=0$ case corresponding to the D'Alembert
equation  in the same way, and   the only change in  formulas (106) would
be the absence of the $m^2$ contribution.

We have noticed before, (32), that if
$\psi (\vec{x},t)$ is a solution of the pseudodifferential--Schr\"{o}dinger
equation  $i\partial _t\psi = [\sqrt {-\triangle + m^2} -m]\psi $, then
$\tilde{\psi }(\vec{x},t) = \psi (\vec{x},t)exp(-imt)$ is a positive energy
solution of the free Klein-Gordon equation $(\Box +m^2)\tilde{\psi }
(\vec{x},t)=0$, since we surely have $i\partial _t\tilde{\psi }=
\sqrt {-\triangle  +m^2}\tilde {\psi }$.
It is clear that the time adjoint  Schr\"{o}dinger equation refers to
negative energy solutions  of the Klein-Gordon equation.
Notice  that we need both positive and negative energy solutions to create
 (upon normalization) a probability density
 $\rho (\vec{x},t)=\psi (\vec{x},t,)\overline {\psi }(\vec{x},t)$.

{\bf Remark 5}: At this point it is useful to emphasize that the time-like
component $j^0(\vec{x},t)$ of the current $j^{\mu }(\vec{x},t)$ is \it not
\rm a probability density itself; by wrongly \cite{petz,petz1} and per
force
assuming that it generally would be the case, all known paradoxes and
difficulties
underlying the refutation of the Klein-Gordon equation as the proper
relativistic generalization of its nonrelativistic Schr\"{o}dinger partner
are revealed \cite{schwe,hol}.
The positive  energy spectrum is not just correlated with positive
(negative)
values of $j_0(\vec{x},t)$, although one can establish such a correlation
for
the total "charge" $e\int j_0(\vec{x},t)d^3x$ \cite{hol,bjo} by assuming
that
$ej_0(\vec{x},t)$ is interpreted as the "charge" density. Even then,
a clean
partition of the positive and negative energy spectra into sets associated
respectively with particles and antiparticles distinguished by the sign of
the charge density is impossible.

\vskip0.2cm

{\bf Remark 6}: The subject of our considerations is essentially a
probabilistic analysis of relativistic quantum mechanics \cite{drell}.
In view of our previous discussion it is clear that $\rho (\vec{x},t)=
|\psi (\vec{x},t)|^2$ is a  probability density of a well defined
stochastic
process, which is non-Markovian in nonstationary situations.
Then, for a general Borel set $A\subset R^3$ we have defined a measure
$\rho (A,t)$ telling us what is the probability for a jump to have its
size
matching a point $\vec{y}\in A$, at time $t$. We are not inclined to think
that a concrete jump refers to an actual "physical particle" that jumps in
space.
In nonrelativistic quantum mechanics, a standard interpretation of
$\rho (\vec{x},t)(\triangle x)^3$ as a probability to locate a \it
physical
\rm particle in a cube of volume  $(\triangle x)^3$ seems to be
consistent.  On the contrary, in relativistic quantum mechanics,
the notions of position
and localizability  and their relation to any experimental
determination of the physical particle position have been a subject of
vigorous disagreements and no general consensus has been  reached.
An  acceptance of the Newton-Wigner localization in the
configuration-space
approach to relativistic quantum theory implies the general breakdown  of
causality \cite{newt,wig,fle,fle1,heg1,heg2} and inevitably implies
superluminal
effects (instantaneous spreading of a localized wave packet). It is by no
means a surprise if one  carefully looks into the jump process features as
revealed in the present paper.
Let us also mention that the Newton-Wigner localization of mass $m=0$
particles is in general impossible (well  known exceptions are massless
particles of spin $0$ and massless spin $1/2$ particles possessing two
helicity states) , while we know how to assign the probability density
notion, and hence a probability measure  $\rho (A,t)$ to a class of
solutions
of the D'Alembert  equation, see e.g. the  arguments of Section 3, albeit
possibly with no  connection to any \cite{jad,janc} "position operator"
notion.

\vskip0.2cm

Each \it scalar \rm positive energy solution $\phi (\vec{x},t)$ of the
free Klein-Gordon equation (105) can be represented
\cite{schwe}  in the manifestly Lorentz covariant form:
$${\phi (\vec{x},t)={\sqrt {2}\over {(2\pi )^{3/2}}} \int d^4k\,
e^{(-ik_{\mu }x^{\mu })}\, \delta (k_{\mu }k^{\mu } - m^2) \Theta (k_0)\,
\Phi (k_0,\vec{k})}\eqno (107)$$
where $k:=(k_0,\vec{k})$, $k_{\mu }k^{\mu }:=k_0^2 - \vec{k}^2$,
$\Phi (k)$
is a scalar and
$\Theta (k_0)$ is the Heaviside function equal to $1$ if $k_0>0$ and
to $0$ otherwise.
The representation (107) extends to all solutions of
$i\partial _t\tilde{\psi }=
\sqrt {-\triangle +m^2}\tilde{\psi }$, and upon changing
$k_0\rightarrow -k_0$
in $\Theta (k_0)$ followed by a complex conjugation of (108), to solutions
of the time adjoint equation as well.
It implies that general solutions of those
pseudodifferential--Schr\"{o}dinger
equations form Lorentz invariant subspaces in the linear space  of all
solutions to the free Klein-Gordon equation.

However, we can not directly infer from the above facts any information
about
how a given pseudodifferential--Schr\"{o}dinger stochastic jump process is
perceived by different relativistic observers. For example the \it
normalization \rm of
Schr\"{o}dinger wave functions is not a relativistically covariant notion.

At this point we adopt the standard definition of the Klein-Gordon scalar
product \cite{schwe}:
$${(\phi _1,\phi _2):= \int_{R^3} d^3x\, [\overline {\phi _1}
\sqrt {-\triangle +m^2}\phi _2\, +\, (\sqrt {-\triangle +m^2}
\overline {\phi _1})\phi _2]}\eqno (108)$$
which is independent of the specific space-like surface of integration.
Both positive and  negative energy solutions are covered in this
definition, albeit separately, with no superposition.  The integrand  in
(108)
should be compared with the time-like component $j^0(x)$ of the conserved
four-current $j^{\mu }(x)$, Eq.(106).

Since the Newton-Wigner position operator
$${[\hat{\vec{x}}_{NW}\phi ](x) \, \equiv  \, i[\nabla _{\vec{k}} -
{{\vec{k}}\over {2(\vec{k}^2+m^2)}}\Phi ](k_0,\vec{k})}\eqno (109)$$
$k_0=\sqrt {\vec{k}^2+m^2}$, $j=1,2,3$, see (107), is Hermitean with
respect
to the scalar product (108), one can introduce a covariant \cite{fle}
localization notion for all positive energy solutions of the free
Klein-Gordon
(and hence pseudodifferential--Schr\"{o}dinger) equation \cite{schwe,ang}.
Indeed, given a positive energy solution of the free Klein-Gordon equation
$\phi (x)$, which we know to solve the pseudodifferential equation as well,
 then, we can introduce a \it  new \rm  solution for both of those equations
as follows \cite{ang}:
$${\phi (x) \rightarrow \phi _{NW}(x):= [(-\triangle +m^2)^{1/4}
\phi ](x)}
\eqno (110)$$
If we take  for granted the Klein-Gordon scalar product normalization,
Eq.(108),of $\phi (x)$, we realize that the common solution of the
Klein-Gordon and pseudodifferential--Schr\"{o}dinger equations:
$ \psi (\vec{x},t):= \phi _{NW}(\vec{x},t)$, may be consistently
normalized according to the Schr\"{o}din\-ger equation rule:
$\rho (x) := |\psi (x)|^2 \Rightarrow  \int_{R^3} d^3x \rho
(\vec{x},t)\, =\, 1$.
As a consequence, and in part because of this normalization, we have
succeeded to associate the  previously investigated stochastic jump
process with the
Newton--Wigner localization. Clearly, we deal with the probability measure
\it identifying \rm the probability that the Newton--Wigner "particle"
can be found at time $t$ in the spatial volume $A$, with the probability of
spatial jumps bounded by this volume, at time $t$:
$${Prob[\vec{X}(t)\in A] = \int_A \rho (\vec{x},t)d^3x =
\int_A |[(-\triangle +m^2)^{1/4}\phi ](\vec{x},t)|^2 d^3x}\eqno (111)$$

The inhomogeneous orthochronous Lorentz mapping  $x'=\Lambda x+a$
($x'^{\mu }= \Lambda ^{\mu } _{\nu }\, x^{\nu } + a^{\mu },
\Lambda ^0 _0>1$) can be
associated with the scalar transformation rule for Klein-Gordon wave
functions
${\phi '(x'_0,\vec{x}')\, =\, \phi (x_0,\vec{x})}$
$\Rightarrow $
$\phi '(x)= \phi (\Lambda ^{-1}(x-a))$.
The transformation acts invariantly in positive and negative energy
subspaces
of solutions, respectively. This fact  implies an extension to
 pseudodifferential-Schr\"{o}dinger equations of motion.

The Klein-Gordon equation is form invariant:
$(\Box ' + m^2)\phi '(x')=(\Box + m^2)\phi (x)=0$,
which allows us to associate with $\phi '(x')$, while normalized
according
to (108), a pseudodifferential-Schr\"{o}dinger stochastic process
according to (110):
$$\tilde {\psi }'(x'):= [(-\triangle ' +m^2)\phi ']^{1/4}(x')$$
$$i\partial _{t'}\tilde {\psi }'(\vec{x}',t') = \sqrt {-\triangle '
+ m^2}
\tilde {\psi }'(\vec{x}',t')$$
$${\Downarrow }\eqno (112)$$
$$i\partial _{t'}\psi '(\vec{x}',t') = [\sqrt {-\triangle ' + m^2} - m]
\psi '(\vec{x}',t')$$
$$\psi '(\vec{x}',t'):=exp(imt')\tilde {\psi}'(\vec{x}',t')$$

The transfer of data about the pseudodifferential-Schr\"{o}dinger  process
from one inertial observer to another is completely determined by Eq.(112).

Now, let us consider the first of equations (87) for the  choice
$H=\sqrt {-\triangle +m^2}$ of the Hamiltonian:
$[i\partial _t -(V-m)]\psi =
\sqrt {-\triangle +m^2}\psi $, where we can regard a general potential
$V-m$
as  a time-like component of a four-vector. In case of electromagnetic
interactions, the presence of the $(-m)$ term  can even be attributed to a
gauge
transformation $A_{\mu }\rightarrow A_{\mu } +\partial _{\mu }\chi $ with
$\chi =-mt$ and $\vec{A}\equiv 0$.
In particular, the relativistic stability of matter studies \cite{lieb} of
the existence of
bound states for the pseudodifferential--Schr\"{o}dinger Hamiltonians,
involve
the Coulomb static potential $eA_0(x)=-(Ze^2/r),\, r:=\sqrt {\vec{x}^2}$.
(We recall that pseudodifferential
Hamiltonian  spectral problems are not limited to the electrostatic
potential
only, but their range of applicability is much wider \cite{carm1}.)

Complex conjugation converts the forward equation into its time
adjoint
and it is clear that $[-i\partial _t - (V-m)]\, \overline {\psi }=
\sqrt {-\triangle +m^2} \,  \overline {\psi }$ holds true.
However, \it no \rm  immediate connection with the general form of the
Klein-Gordon equation in the presence of electromagnetic interactions
(charge $e$ particles), $(i\partial _{\mu } -eA_{\mu })
(i\partial ^{\mu }-
eA^{\mu })\phi (x)=m^2\phi (x)$, can be established, in general.

On the other hand,  a pedestrian intuition  behind the associated
notion of a
\it relativistic atom \rm is quite helpful for a deeper understanding
of the particular r\^{o}le of the Lorentz
covariance of the Klein-Gordon and D'Alembert equations in the context of
\it free \rm pseudodifferential Schr\"{o}dinger equations.
Namely,  the atom itself is always considered to be at rest, with
the frame of reference attached to the nucleus, which in turn is a
source of an electrostatic field.
In this particular frame of reference the
pseudodifferential--Schr\"{o}dinger equation
$[i\partial _t - (V-m)]\psi =\sqrt {-\triangle +m^2}\psi $
is defined.
The same pertains to the general pseudodifferential--Schr\"{o}dinger
problem
with a minimal electromagnetic coupling due to external fields
$${[i\partial _{t} - (eA^0-m)]\psi (\vec{x},t) =
\sqrt {\Sigma _{j=1}^3 [(i\nabla _j- eA_j)(i\nabla ^j-eA^j)] +m^2} \,
\psi (\vec{x},t)}\eqno (113)$$
which is a frame-of-reference-dependent notion, see e.g. \cite{such}.

{\bf Remark 7}: A Euclidean version of (113) was investigated in
\cite{ichi1}
and an explicit construction was given of the Feynman-Kac path integral
formula for the corresponding semigroup kernel. It  involves paths, and
conditional measures over paths, of a time homogeneous L\'{e}vy process.

{\bf Remark 8}: The problem of how a stochastic process can be
perceived by
different relativistic observers has been considered before \cite{garb2}
in connection with  certain  (Markov) rotational diffusions on an $S_3$
manifold ($SU(2)\times SU(2)$ case specialized to spin $1/2$), with the
Euler angles parametrization established relative to a fixed
three-dimensional orthonormal basis.
\vskip0.2cm
Our discussion was confined to  the mass $m>0$ case, but in view of the
general non-existence of the Newton-Wigner localization in the mass $m=0$
case, an extension of previous  relativistic covariance arguments
needs some care.
The present localization is known to be admissible for massless spin $0$
particles. As well, we do not literally need the Newton-Wigner-like
position
operator and the associated notion of localization at a spatial point to
invoke a  substantial part of the previous arguments.
(In fact, a maximal localizability of photons on a circle, hence in a
subset of $R^2$, was established in \cite{jad}.)
As long as (108) is extended to mass $m=0$ particles as the normalization
condition for wave functions, and the Borel set $A$ is never point-wise,
we can safely go through (110)-(112).

{\bf Remark 9}: There have been numerous attempts to associate  the
Klein-Gordon
equation with stochastic processes. In addition to \cite{ang} let us
mention
a number of other attempts \cite{gue,lehr,vig,zast,mar,ser,mor,ja,garb1}.
None of them  can be viewed as a "derivation" of the Klein-Gordon field
from certain "first" stochastic principles.
Except for \cite{ang}, all these attempts  exploited a formal similarity of
Eqs.(106) to local conservation laws shared by nonrelativistic Markov
diffusions \cite{blanch} and  to analogous laws in relativistic kinetic
theory \cite{garb1,isr,goto}.
The status of the Markov property has been found disputable, since its
violation is implicit if the relativistic invariance of diffusion
(Kolmogorov)
equations in Minkowski space is required \cite{lop,sch,ha,dud}.
In the present paper, we have found the Markov property admissible
only in case
of the measure preserving (stationary case) stochastic jump dynamics;
see e.g. Section 2.
\vskip0.2cm
{\bf Acknowledgement}: We express our  gratitude to Professor G. G. Emch
for a helpful discussion.

\end
{document}